# Experimental Observation of Bethe Strings

Zhe Wang[1], Jianda Wu[2], Wang Yang[2], Anup Kumar Bera[3,4], Dmytro Kamenskyi[5], A.T.M. Nazmul Islam[3], Shenglong Xu[2], Joseph Matthew Law[6], Bella Lake[3,7], Congjun Wu[2], Alois Loidl[1]

[1]Experimental Physics V, Center for Electronic Correlations and Magnetism, Institute of Physics, University of Augsburg, 86135 Augsburg, Germany
[2]Department of Physics, University of California, San Diego, California 92093, USA
[3]Helmholtz-Zentrum Berlin für Materialien und Energie, 14109 Berlin, Germany
[4]Solid State Physics Division, Bhabha Atomic Research Centre, Mumbai 400085, India
[5]High Field Magnet Laboratory, Radboud University, 6525 ED Nijmegen, The Netherlands
[6]Hochfeld Magnetlab Dresden, Helmholtz-Zentrum Dresden-Rossendorf, 01314 Dresden, Germany
[7]Institut für Festkörperphysik, Technische Universität Berlin, 10623 Berlin, Germany

**Abstract**:

Almost one century ago, string states - complex bound states (Wellenkomplexe) of magnetic excitations - have been predicted to exist in one-dimensional quantum magnets and since then become a subject of intensive theoretical study. However, experimental realization and identification of string states in condensed-matter systems remains an unsolved challenge up to date. Here we use high-resolution terahertz spectroscopy to identify string states in the antiferromagnetic Heisenberg-Ising chain $SrCo_2V_2O_8$ in strong longitudinal magnetic fields. We observe complex bound states (strings) and fractional magnetic excitations (psinons and antipsinons) in the field-induced critical regime, which are precisely described by the Bethe ansatz. Our study reveals that two-string and three-string states govern the quantum spin dynamics close to the quantum criticality, while the fractional excitations are dominant at low energies, reflecting the antiferromagnetic quantum fluctuations.

**Introduction**:

The one-dimensional spin-1/2 Heisenberg model is the paradigmatic model of interacting spin systems. Enlightening insight into fundamental aspects of magnetism can be provided by studying the spin dynamics of this quantum model. Early in 1930s Bethe has introduced a systematic ansatz to calculate its exact eigenvalues and eigenstates (*1*). Among the generic solutions of spin waves, Bethe also predicted bound states of spin waves. Spin waves in the scattering states, characterized by real momenta, collide and propagate throughout the chain, while the bound states – described by complex momenta – correspond to strings of flipped spins which move as entities along the chain (*1,2,3*). The Bethe ansatz can be applied in a far more general context to study a variety of one-dimensional integrable problems (*4*), such as the fermionic Hubbard model (*5*) and the non-equilibrium behavior of bosonic atom chains in optical lattices (*6,7*). Also in the context of quantum communication (*8*), the bound states can play an important role in the transport properties across a channel of qubits (*9*). Recently this powerful approach has been even extended to study the string dynamics in the $N = 4$ super Yang-Mills theory (*10*), where the string, as a successful concept in string theory, can be represented by a chain of identical scalar fields, and the string excitations correspond to inserting different scalar fields (oscillators) into the chain (*10,11*). With $n$ identical oscillators inserted the so-called Berenstein-Maldacena-Nastase states (*11*), as counterparts of string excitations on the light-cone, can be mapped onto the bound states with $n$ complex momenta ($n$-string states) in the one-dimensional Heisenberg model (*10*). Hence, the study of quantum spin dynamics in the Heisenberg model also provides a new perspective for the string theory. Here we present the results of high-resolution terahertz spectroscopy on SrCo$_2$V$_2$O$_8$ in longitudinal magnetic fields, identifying unambiguously string excitations in a condensed-matter system.

The realization of the 1D Heisenberg-Ising model in SrCo$_2$V$_2$O$_8$ relies on its crystal structure and dominant magnetic interactions (Fig. 1A). The screw chains of CoO$_4$ tetrahedra with fourfold screw axis running along the crystallographic *c* direction are arranged in a tetragonal structure. Due to spin-orbit coupling, the atomic magnetic moments of the Co$^{2+}$ ions, comprising spin and orbital degrees of freedom, are exposed to an Ising anisotropy. The crystal-electric field in the CoO$_4$ tetrahedra lifts the twelvefold degeneracy of the Co$^{2+}$ moments, resulting in a Kramers-doublet ground state with the total angular momentum 1/2 (*12*). Magnetization and neutron diffraction experiments reveal that the Ising anisotropy forces the atomic magnetic moments along the *c* axis, with a Néel-type collinear antiferromagnetic order stabilized below $T_N = 5$ K (*13*).

Superexchange interactions between the magnetic moments in SrCo$_2$V$_2$O$_8$ are described by the Hamiltonian of the one-dimensional spin-1/2 Heisenberg-Ising model,

$$H = J \sum_{n=1}^{N} \left[ \left( S_n^x S_{n+1}^x + S_n^y S_{n+1}^y \right) + \Delta S_n^z S_{n+1}^z \right] - g_\parallel \mu_B B \sum_{n=1}^{N} S_n^z \quad (1)$$

where $J > 0$ is the antiferromagnetic coupling between neighboring spins and $\Delta > 1$ takes into account the Ising anisotropy between the longitudinal and transverse spin couplings. The last term includes the Zeeman interaction in a longitudinal magnetic field $B$ along the *c* axis with the *g*-factor $g_\parallel$ and the Bohr magneton $\mu_B$.

The Néel ground state at zero field can be illustrated by an antiparallel alignment of neighboring spins, corresponding to the total spin-*z* quantum number $S_T^z = 0$ (Fig. 1A). One spin-flip creates an excitation of two spinons (*14*) that can separately propagate along the chain by subsequent spin-flips. In momentum space they form a two-particle excitation continuum (*15,16,17,18*) that is gapped above the ground state – the spinon vacuum (*19*).

Via the Zeeman interaction a longitudinal magnetic field can reduce and finally close the spin gap at a critical field $B_c$ (*20*). Before reaching the fully polarized state ($B > B_s$, Fig. 1A) where the elementary excitations are gapped magnons, the system enters a gapless phase corresponding to the critical regime ($B_c < B < B_s$). In this regime, a general ground state with an arbitrary value of $S_T^z$ is illustrated in Fig. 1B, and fundamentally new states can be excited by flipping a single spin (Figs. 1C to 1F). According to the Bethe ansatz (*1*), the spin excitations in the critical regime can be Bethe strings of length $n$ ($n$-string states) (*2,3*) corresponding to $n$-complex momenta, or pairs of psinon-psinon/psinon-antipsinon (*19,21*) with real momenta (*22,23,24*). Psinons and antipsinons, introduced by Karbach and Müller (*19,21*), are elementary fractional magnetic excitations. Compared to spinons in zero field (*15*), in the critical regime psinons/antipsinons also form continua of two-particle (one-pair) or multi-pair excitations but with different dispersions (*19,25*). The psinon-(anti)psinon continuum is gapless at certain momenta that follow characteristic dependencies on the longitudinal magnetic field (*24*). The string states, depending on their length $n$, exhibit distinctive excitation continua, which are gapped and located at higher energies (*24*).

Since excitations with $\Delta S_T^z = +1$ or $-1$ are allowed by selection rules that govern the interaction with the photon magnetic field or the neutron magnetic moment, the excited states illustrated in Figs. 1C to 1F should be observable by optical or neutron scattering experiments. Here, we report terahertz optical spectroscopy on the Heisenberg-Ising antiferromagnetic chain $SrCo_2V_2O_8$ in a longitudinal magnetic field up to 30 T. Our study provides clear experimental evidence on the high-energy quantum excitations of 2-string and 3-string states, as well as on the low-lying fractional excitations of psinons/antipsinons, characterizing the quantum spin dynamics of the one-dimensional spin-1/2 Heisenberg-Ising model.

In a longitudinal magnetic field, $S_T^z$ is a good quantum number. Hence, the eigenstates of the Heisenberg-Ising model can be accordingly classified, and described by a general Bethe-ansatz wavefunction

$$|\varphi\rangle = \sum_{1 \leq n_1 \leq \cdots \leq n_r \leq N} a(n_1, n_2, \ldots, n_r) |n_1, n_2, \ldots, n_r\rangle$$

for the total spin-$z$ quantum number $S_T^z = N/2 - r$ (Fig. 1B), with $N$ denoting the length of the chain, and $r$ flipped spins at the sites of $n_1, n_2, \ldots$, and $n_r$ in the chain with respect to the fully polarized state $|\ldots \rightarrow\rightarrow\rightarrow \cdots \rangle$, i.e. $|n_1, n_2, \ldots, n_r\rangle = S_{n_1}^- S_{n_2}^- \ldots S_{n_r}^- |\ldots \rightarrow\rightarrow\rightarrow \cdots \rangle$ where $S_{n_j}^\pm \equiv S_{n_j}^x \pm i S_{n_j}^y$ is the operator flipping the spin of the $n_j$ site. We use the Bethe ansatz to obtain the coefficients $a(n_1, n_2, \ldots, n_r)$ and eigenenergies of the ground state as well as excited states for every $S_T^z$ sector (*24*). Corresponding to the real and complex momenta in the solutions of the Bethe-ansatz equations (*1*), the excited states of psinon-(anti)psinons and $n$-strings are labelled as $R_q$ and $\chi_q^{(n)}$, respectively, with the subscript indexing the corresponding transfer momentum. Those excitations allowed in optical experiments obey the selection rule $\Delta S_T^z = +1$ or $-1$, which contribute to the dynamic structure factor $S^{-+}(q,\omega)$ or $S^{+-}(q,\omega)$, respectively. The two quantities are defined by $S^{a\bar{a}}(q,\omega) = \pi \sum_\mu |\langle \mu | S_q^{\bar{a}} | G \rangle|^2 \delta(\omega - E_\mu + E_G)$ in which $S_q^\pm = \frac{1}{\sqrt{N}} \sum_n e^{iqn} S_n^\pm$ and $\bar{a} = -a$ with $a = +$ or $-$,

and $|G\rangle$ and $|\mu\rangle$ are the ground state and the excited state with eigenenergies of $E_G$ and $E_\mu$, respectively (*24,26*). Hence, we can quantitatively single out contributions of string excitations, as well as of psinon-(anti)psinon pairs, to the relevant dynamic structure factors. The string excitations with higher energies and characteristic field dependencies can be well distinguished from the low-energy psinon-(anti)psinon pairs. This enables us to precisely compare to the experimental results as a function of longitudinal field and to identify the nature of each observed mode.

**Results and Discussion:**

Figure 2A shows transmission spectra at various frequencies below 1 THz as a function of longitudinal magnetic fields. At 0.195 THz, two transmission minima can be observed at 2.41T (mode $1-$) and 6.18 T (mode $R_0$), below and above the critical field $B_c = 4$ T, respectively. On increasing magnetic fields, mode $1-$ shifts to lower, and mode $R_0$ to higher frequencies. The mode $1-$ together with the modes $1+$, $2-$, $2+$, and $3-$ observed at higher frequencies (0.39 and 0.59 THz) are known as confined spinon excitations (*27*) due to the inter-chain couplings in the gapped Néel-ordered phase ($B < B_c$). Well below the critical field (Fig. 2B), the confined spinons exhibit Zeeman splitting with the linear field dependence (*27*). Close to the critical field $B_c$, the mode $1-$ strongly softens, concomitant with significant hardening of the mode $1+$. This indicates that the inter-chain couplings are suppressed above $B_c$, and the system enters the field induced critical regime of one dimension. Completely different excitation spectra are observed in the critical regime. In the same frequency range (Fig. 2A), we unambiguously observe three sharp modes as denoted by $R_0$, $\chi_0^{(2)}$, and $\chi_\pi^{(2)}$. With increasing magnetic field well above the critical field, the eigenenergies of the three modes increase linearly with different slopes (Fig. 2B).

Using magneto-optic spectrometry in a high-field laboratory, we are able to extend the search for magnetic excitations to a larger spectral range and to higher magnetic fields up to 30 T, for the whole critical regime ($B_c < B < B_s$) and the field-polarized ferromagnetic phase ($B > B_s = 28.9$T). As displayed in Fig. 3, the magnetic excitations are represented by the peaks in the absorption-coefficient spectra at various magnetic fields. At 10 T we not only identify the modes $R_0$, $\chi_0^{(2)}$, and $\chi_\pi^{(2)}$ at 1.66, 3.67, and 4.23 meV, respectively, in a sequence of increasing energies, but also a higher-energy mode $R_{\pi/2}$ at 5.82 meV. While the modes $R_0$ and $\chi_\pi^{(2)}$ have comparable absorption coefficients, the mode $\chi_0^{(2)}$ is much weaker, which is consistent with the low-field measurements (e.g. 0.59 THz spectrum in Fig. 2A) and the features expected from the Bethe-ansatz results (*24*). In higher magnetic fields, the mode $R_{\pi/2}$ softens, while the other modes $R_0$, $\chi_0^{(2)}$, and $\chi_\pi^{(2)}$ shift to higher energies. Above the low-energy phonon bands (*24*), we resolve a fifth and high-energy magnetic excitation $\chi_{\pi/2}^{(3)}$ of 14.6 meV at 13 T, which shifts to higher energy with increase of the longitudinal field (Fig. 3B). The field dependence of the eigenfrequencies of the five observed modes $R_0$, $R_{\pi/2}$, $\chi_0^{(2)}$, and $\chi_\pi^{(2)}$, and $\chi_{\pi/2}^{(3)}$ is summarized in Fig. 4 (symbols). The field dependencies are linear for all the five modes, each with distinct slope and characteristic energy.

From the Bethe-ansatz calculations, we can single out various excitations and evaluate their respective contributions to the dynamic structure factors $S^{+-}(q,\omega)$ and $S^{-+}(q,\omega)$ (*22,23,24*). In Figure 4 the peak frequencies in $S^{+-}(q,\omega)$ and $S^{-+}(q,\omega)$ are shown as a function of magnetic field (solid lines) for various values of the transfer momenta $q = 0$, $\pi/2$, and $\pi$, in accord with Brillouin-zone folding due to the fourfold screw-axis symmetry of the spin chain (*24*). Excellent agreement between theory and experiment is achieved for all the five distinct magnetic excitations $R_0$, $R_{\pi/2}$, $\chi_0^{(2)}$, and $\chi_\pi^{(2)}$, and $\chi_{\pi/2}^{(3)}$, using the exchange interaction $J = 3.55$ meV, the $g$-factor $g_\| = 6.2$, and the Ising anisotropy $\Delta = 2$ (*24*). The precise comparison allows to unambiguously identify the nature of the observed excitations: $R_0$ characterizes psinon-antipsinon pairs at $q = 0$, while $R_{\pi/2}$ are psinon-psinon pairs at $q = \pi/2$. The psinon-antipsinon excitations $R_0$, related to the single spin-flip operator $S^-_{q=0}$, evolve from the magnon mode $M_0$ in the field-polarized ferromagnetic phase, where the largest absorption is observed experimentally (Figs. 3A and 4). Most strikingly, we are able to detect and determine the 2-string states $\chi_0^{(2)}$ and $\chi_\pi^{(2)}$ at $q = 0$ and $q = \pi$, respectively, and even the 3-string states $\chi_{\pi/2}^{(3)}$ for $q = \pi/2$.

The branch of psinon-psinon pairs $R_{\pi/2}$ belongs to $S^{-+}(q,\omega)$ and obeys the selection rule of $\Delta S_T^z = +1$. Hence, the psinon-psinon excitations correspond to flipping one spin to the magnetic field direction (Fig. 1C), which will decrease the Zeeman energy. Thus the mode $R_{\pi/2}$ softens with increasing field. In contrast, the strings and psinon-antipsinon pairs correspond to $S^{+-}(q,\omega)$, where one spin is flipped antiparallel to the magnetic field ($\Delta S_T^z = -1$, Figs. 1D to 1F), thus their energies increase in magnetic field. The linear dependencies, essentially arising from the linear dependence of the Zeeman energy on magnetic field, are significantly renormalized due to one-dimensional many-body interactions.

Bethe-ansatz analysis also shows that the contributions of different excitations to the spin dynamics are strongly field-dependent (*22,23,24*). Close to the fully field-polarized limit, psinon-antipsinon pairs dominate $S^{+-}(q,\omega)$. Their contributions decrease significantly with decreasing magnetic field, while at the same time string excitations become more and more important (*24*). Especially, below the half-saturated magnetization, the 2-string and 3-string states contribute substantially and finally govern the spin dynamics close to the critical field (*24*). This is indeed observed in the experiment: with decreasing magnetic field, the mode $R_0$ rapidly becomes weaker, while the 2-string states exhibit a slightly increased absorption.

The Bethe strings revealed here in the quantum critical spin chain represent an outstanding example of experimental realization of strongly correlated quantum-states in condensed-matter systems (*28,29*). Our results pave the way to study complex magnetic many-body states and their dynamics in quantum magnets (*30*). They may also shed light on the study of quantum dynamics in quasi-one dimensional ultra-cold atom systems (*7*).

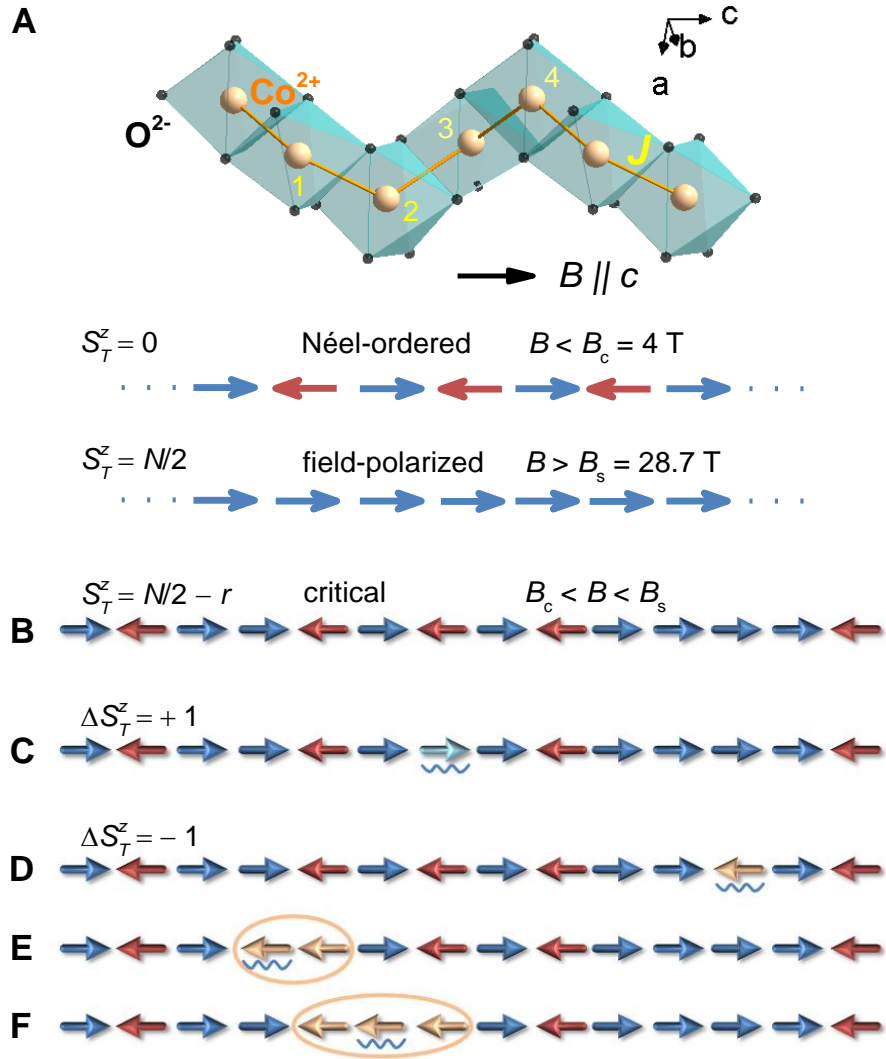

**Fig. 1. Quantum spin chain in SrCo$_2$V$_2$O$_8$ and characteristic magnetic excitations of one dimension in the critical regime: psinons-(anti)psinons and strings.**

(**A**) Chain structure of SrCo$_2$V$_2$O$_8$ with a four-fold screw axis along the *c* direction: at 1.7 K Néel-ordered and field-polarized states are stabilized for longitudinal magnetic fields $B < B_c = 4$ T and $B > B_s = 28.7$ T, respectively. (**B**) A representative configuration of the ground state in the critical regime ($B_c < B < B_s$) for the total spin-z quantum number $S_T^z = N/2 - r$ with $r$ flipped spins with respect to the fully polarized state. Excitation states are allowed by the selection rule $\Delta S_T^z = +1$ for psinon-psinon (**C**), or by $\Delta S_T^z = -1$ for psinon-antipsinon, 2-string, and 3-string (**D** to **F**), that govern the interaction with the magnetic field of a photon. While the psinon-(anti)psinons can propagate throughout the chain without forming bound states, the 2-strings and 3-strings move only as entities in the chain. The flipped spin with respect to the ground state (**B**) is indicated by a wiggly line for every excited state (**C** to **F**).

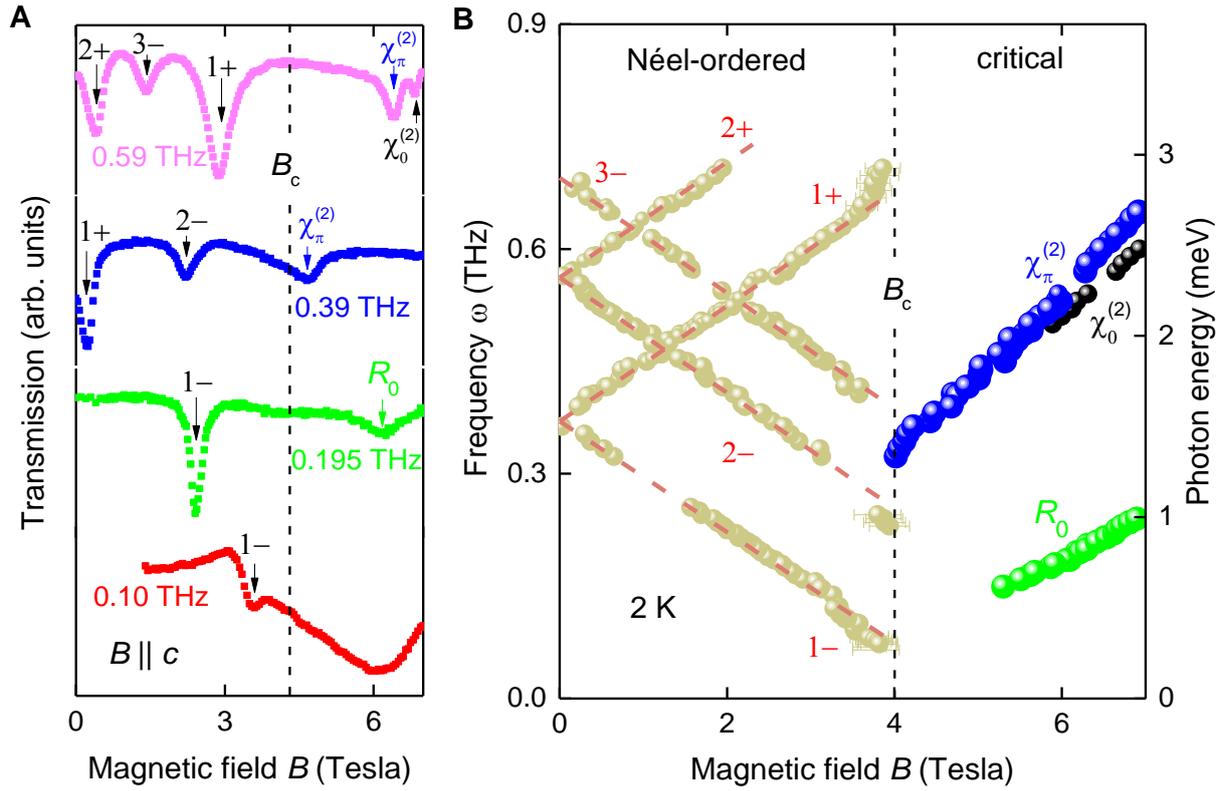

**Fig. 2. Softening of spinons and emergent magnetic excitations at the quantum phase transition in SrCo$_2$V$_2$O$_8$.**

(**A**) Transmission spectra of magnetic excitations in SrCo$_2$V$_2$O$_8$ for various frequencies below 1 THz, measured with the applied longitudinal magnetic field $B \parallel c$ and the electromagnetic wave propagating along the $c$ axis. Magnetic resonance excitations corresponding to transmission minima are indicated by arrows. The field-induced phase transition from the Néel-ordered phase to the critical regime is indicated by the vertical dashed line at the critical field $B_c = 4$ T.

(**B**) Eigenfrequencies of the resonance modes as a function of the applied magnetic field. For $B < B_c$, a series of confined fractional spinon excitations split in low fields (1±, 2±, etc) and follow linear field dependency (*27*). Above $B_c$, new modes $R_0$, $\chi_0^{(2)}$, and $\chi_\pi^{(2)}$ emerge with completely different field dependencies. Clear deviations from the linear field-dependence appear when approaching the critical field. Error bars indicate the resonance linewidths.

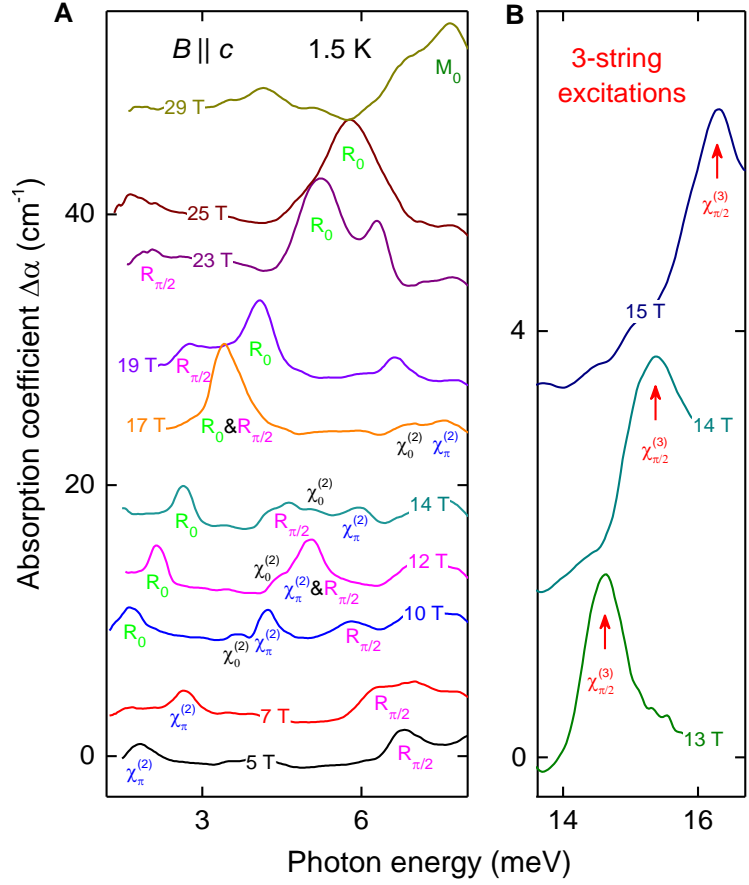

**Fig. 3. Absorption spectra of psinon-psinon, psinon-antipsinon, 2-string, and 3-string excitations for $B_c < B < B_s$, and of magnon for $B > B_s$ in SrCo$_2$V$_2$O$_8$.** Absorption spectra for various longitudinal magnetic fields in the critical regime in (**A**) the low-energy and (**B**) the high-energy spectral range. (**A**) Four types of excitations, $R_0$, $R_{\pi/2}$, $\chi_0^{(2)}$, and $\chi_\pi^{(2)}$, each with a characteristic field dependence are observed in the critical regime ($B_c < B < B_s$), and the mode $R_0$ evolves from the mode $M_0$ above $B_s$. While the mode $R_{\pi/2}$ softens with increasing fields, eigenenergies of the other modes increase. (**B**) A higher-energy mode $\chi_{\pi/2}^{(3)}$ can be resolved at relatively low magnetic fields. The spectra are shifted upwards proportional to the corresponding magnetic fields.

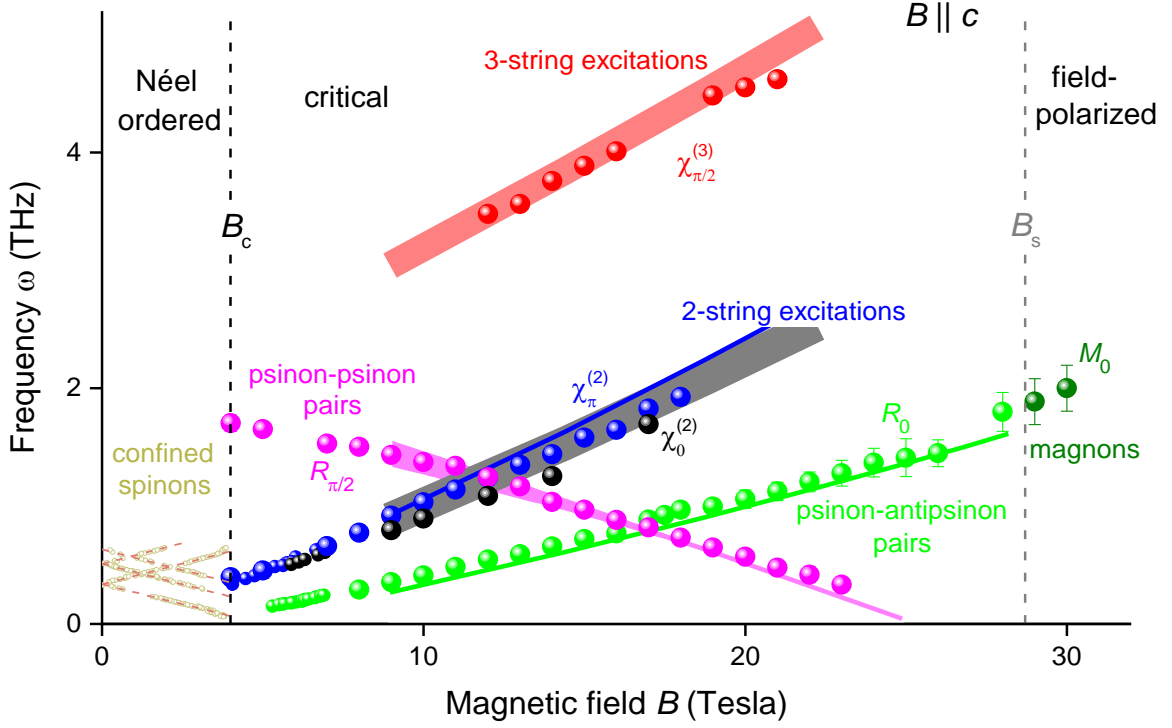

**Fig. 4. Quasiparticle excitations in the longitudinal-field Heisenberg-Ising chain SrCo₂V₂O₈: experiment and theory.**

Eigenfrequencies as a function of longitudinal magnetic field for all magnetic modes observed experimentally are shown by symbols. Below $B_c$ confined spinons are observed in the Néel-ordered phase. In the critical regime ($B_c < B < B_s$), quasiparticles of psinon-psinon pairs ($R_{\pi/2}$ at $q = \pi/2$), psinon-antipsinon pairs ($R_0$ at $q = 0$), and of complex bound states 2-strings ($\chi_\pi^{(2)}$, $\chi_0^{(2)}$ at $q = \pi, 0$) and 3-strings ($\chi_{\pi/2}^{(3)}$ at $q = \pi/2$) are identified by the field dependencies of their eigenfrequencies. Above $B_s$, magnons ($M_0$ at $q = 0$) are observed in the field-polarized ferromagnetic phase.

Solid lines display the results of the dynamic structure factors $S^{-+}(q, \omega)$ and $S^{+-}(q, \omega)$ at $q = 0, \pi/2$, and $\pi$ for the corresponding quasiparticles of the one-dimensional spin-1/2 antiferromagnetic Heisenberg-Ising model in Eq. 1. For all the five modes ($R_0$, $R_{\pi/2}$, $\chi_0^{(2)}$, and $\chi_\pi^{(2)}$, and $\chi_{\pi/2}^{(3)}$), excellent agreement between theory and experiment is achieved with the exchange interaction $J = 3.55$ meV, the $g$-factor $g_\parallel = 6.2$, and the Ising anisotropy $\Delta = 2$. Experimental and theoretical linewidths are indicated by error bars and shaded areas, respectively (*24*).

**Acknowledgements:**

We thank Immanuel Bloch, Michael Karbach, and Xenophon Zotos for useful discussions. We acknowledge partial support by the Deutsche Forschungsgemeinschaft via Transregio 80, and also the support of the HFML-RU/FOM and the HLD at HZDR, members of the European Magnetic Field Laboratory (EMFL). J.W, W.Y., S.L.X., and C.W. are supported by the NSF Grant No. DMR-1410375 and AFOSR Grant No. FA9550-14-1-0168. C.W. acknowledges the support by the President's Research Catalyst Awards CA-15-327861 from the University of California Office of the President.

# Supplementary Materials:

# Experimental Observation of Bethe Strings

Zhe Wang[1],* Jianda Wu[2], Wang Yang[2], Anup Kumar Bera[3,4], Dmytro Kamenskyi[5], A.T.M. Nazmul Islam[3], Shenglong Xu[2], Joseph Matthew Law[6], Bella Lake[3,7], Congjun Wu[2], Alois Loidl[1]

[1]Experimental Physics V, Center for Electronic Correlations and Magnetism, Institute of Physics, University of Augsburg, 86135 Augsburg, Germany

[2]Department of Physics, University of California, San Diego, California 92093, USA

[3]Helmholtz-Zentrum Berlin für Materialien und Energie, 14109 Berlin, Germany

[4]Solid State Physics Division, Bhabha Atomic Research Centre, Mumbai 400085, India

[5]High Field Magnet Laboratory, Radboud University, 6525 ED Nijmegen, The Netherlands

[6]Hochfeld Magnetlab Dresden, Helmholtz-Zentrum Dresden-Rossendorf, 01314 Dresden, Germany

[7]Institut für Festkörperphysik, Technische Universität Berlin, 10623 Berlin, Germany

## Supplementary Materials:

Materials and Methods
Theoretical Background
Figures S1 – S10
References 31 – 33

**Materials and Methods**

**Sample preparation**

High-quality single crystals of $SrCo_2V_2O_8$ were grown using the floating-zone method. Crystal structure and magnetic properties were characterized by X-ray diffraction, neutron diffraction, and magnetization measurements (*13*). For the optical experiments, single crystals were oriented using X-ray Laue diffraction and cut perpendicular to the tetragonal *c* axis with a typical surface area of $4 \times 4$ mm$^2$ and a thickness of 1 mm.

**Terahertz spectroscopy in magnetic fields**

Low-frequency optical experiments were carried out in Augsburg. For the transmission spectroscopy below 1 THz, backward wave oscillators were used as tunable sources of electromagnetic waves. A magneto-optic cryostat (Oxford Instruments/Spectromag) was utilized to apply external magnetic fields up to 7 T and to control temperatures down to 2 K. High-field optical measurements were performed in the High Field Magnet Laboratory in Nijmegen. Transmission spectra were measured using a Fourier-transform spectrometer Bruker IFS-113v, combined with a 30-Tesla Bitter electromagnet. THz electromagnetic waves were generated by a Mercury lamp and were detected by a silicon bolometer. For all the optical measurements, the external magnetic fields were applied parallel to the crystallographic *c* axis (longitudinal field, $\boldsymbol{B} \parallel c$) and parallel to the propagation direction of the electromagnetic wave (Faraday configuration, $\boldsymbol{B} \parallel \boldsymbol{k}$).

**Crystal and magnetic structure of $SrCo_2V_2O_8$**

The spin-1/2 Heisenberg-Ising antiferromagnetic chain system $SrCo_2V_2O_8$ crystallizes in a tetragonal structure with the space group *I*4$_1$*cd* (Fig. S1), and the lattice constants $a = 12.2710(1)$ Å and $c = 8.4192(1)$ Å at room temperature. The screw-chain structure is based on edge-shared $CoO_4$ tetrahedra. The screw axis of every chain is along the crystallographic *c* direction with a period of four $Co^{2+}$ ions (Fig. S1A). In each unit cell, there are four chains, two with the left-handed screw-axes and the other two with the right-handed ones (Fig. S1B). The leading inter-chain couplings are between $Co^{2+}$ ions from the neighboring chains with the same chirality (*31*), as indicated in Fig. S1B. Compared to the intra-chain interaction, the inter-chain coupling is almost negligible, i.e. $J_\perp/J < 10^{-2}$ (*31*).

Due to crystal-field effects and spin-orbit coupling, the atomic magnetic moments of each $Co^{2+}$ ion form a ground state of Kramers doublet corresponding to the total angular momentum of 1/2, which comprises spin and orbital degrees of freedom. According to the crystal-field theory (*12,32*) and electron spin resonance measurements (*27*), the magnetic gap between the two lower-lying Kramers doublets is about 22 meV. The nearest-neighbor exchange interactions between the Kramers doublets in the ground state (corresponding to $m_J = \pm 1/2$) can be described by the spin-1/2 Heisenberg-Ising antiferromagnetic model. Below $T_N \sim 5$ K, a Néel-ordered phase is stabilized in zero magnetic field (*13*).

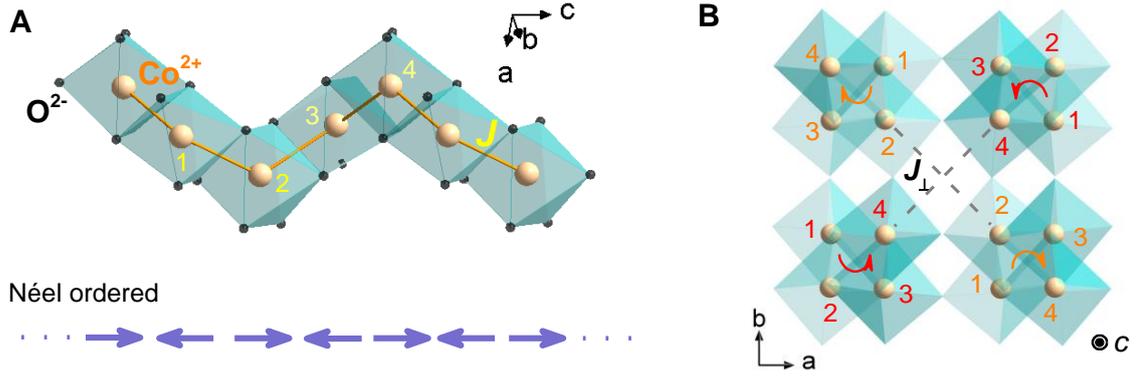

**Fig. S1. Crystal and magnetic structure of SrCo$_2$V$_2$O$_8$.**
(**A**) Screw chain structure is constituted by edge-shared CoO$_4$ tetrahedra. Each chain has screw-axis symmetry with a period of four Co ions (as numbered by the integers 1, 2, 3, 4 in the figure), corresponding to the lattice constant along the *c* axis. The Néel ordered phase is illustrated by antiparallel arrows representing magnetic moments at the Co$^{2+}$ sites. Intra-chain nearest-neighbor interaction is denoted by $J$.
(**B**) Viewing from the *c* axis, each unit cell contains four screw chains with left- or right-handed screw axes, respectively. The leading inter-chain coupling $J_\perp$ is indicated, which is between the Co$^{2+}$ ions in the same layer (denoted by the same integer number of Co site) and from chains with the same chirality. It is very small compared to the intra-chain interaction, i.e. $J_\perp/J < 10^{-2}$ (*31*).

### High-field magnetization in SrCo$_2$V$_2$O$_8$

Magnetization measurements were performed at the Dresden High Magnetic Field Laboratory in a pulsed magnetic field up to 60 T. Fig. S2A shows the magnetization in a longitudinal field $B \parallel c$ at 1.7K. The magnetization curve exhibits distinct features in different phases and a clear signature of a quantum phase transition.

At zero magnetic field, the Néel-ordered phase is characterized by gapped fractional spinon excitations. A clear onset of magnetization occurs at $B_c = 4$ T indicating the phase transition to the 1D gapless phase (critical regime). Slightly above $B_c$, the magnetization increases quasi-linearly, and then strongly nonlinear above 10 T. This is also reflected by the field-derivative of the magnetization $dM/dH$: a nearly constant value is followed by an evident upturn increase in higher fields (Fig. S2B). Field-induced phase transitions at $B_c$ and $B_s$ are evidenced by the anomalies in the field derivative of magnetization. Above $B_s = 28.7$ T, the spins are fully field polarized and the system enters a field-induced ferromagnetic phase with saturated magnetization.

We use the spin-1/2 Heisenberg-Ising antiferromagnetic model (Eq. S1) to simulate the field dependence of the magnetization in the gapless 1D critical regime between $B_c$ and $B_s$ (*20*). Using the same parameters ($J = 3.55$ meV, $g = 6.2$, and $\Delta = 2$) as determined from the excitation spectra (Fig.4), we can describe the experimental magnetization curve quite well.

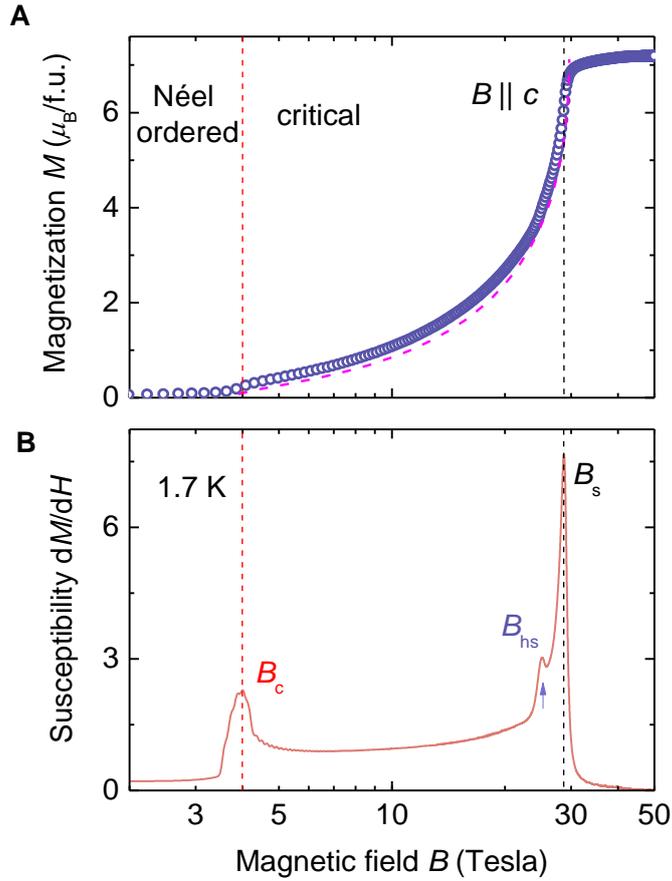

**Fig. S2 High-field magnetization and magnetic susceptibility of SrCo$_2$V$_2$O$_8$.**
(**A**) Magnetization $M$ as a function of an applied longitudinal magnetic field $B$ along the Ising axis ($B \parallel c$) measured at 1.7 K. Theoretical magnetization of the Heisenberg-Ising chain model is shown by the dashed line.
(**B**) Magnetic susceptibility $dM/dH$ as a function of the applied longitudinal field $B$. A quantum phase transition from the Néel-ordered phase to the critical phase is revealed by the onset of magnetization and the peak in the susceptibility curve at the critical field $B_c = 4$ T. Saturated magnetization is observed above the field $B_s = 28.7$ T and indicated by the sharp peak of the susceptibility. A small anomaly at $B_{hs} = 25$ T seen in the susceptibility is close to the field of half-saturated magnetization.

### Low-energy phonon spectrum in SrCo$_2$V$_2$O$_8$

The low-energy phonon reflection spectrum was measured using Fourier transform infrared spectroscopy. Figure S3 shows the phonon spectra of SrCo$_2$V$_2$O$_8$ measured for the polarization $E^\omega \parallel a$ in the relevant spectral range. The strong phonon reflection from 8 to 13.5 meV strongly reduces the transmission at the corresponding spectral range.

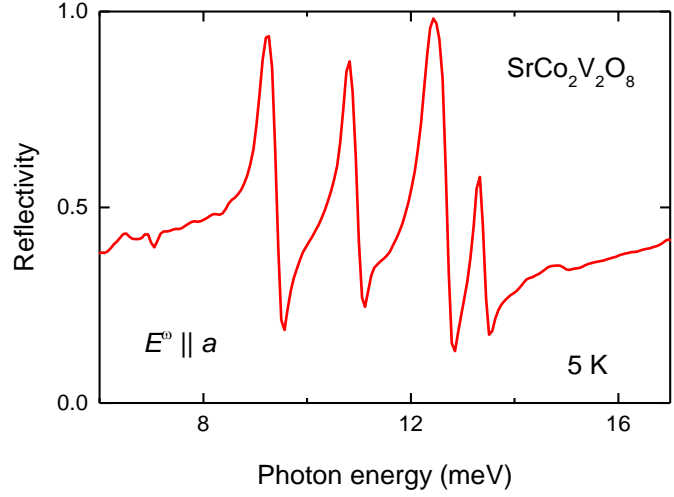

**Fig. S3 Low-energy phonon spectrum of SrCo₂V₂O₈.** The phonon spectra of SrCo$_2$V$_2$O$_8$ measured for the polarization $E^\omega \parallel a$ at 5 K. Strong phonon reflection is observed in the spectral range from 8 to 13.5 meV.

**Theoretical Background**

<u>**Bethe ansatz formalism**</u>

We use Bethe ansatz to obtain the eigenenergies and eigenwavefunctions of the spin-1/2 antiferromagnetic Heisenberg-Ising model

$$H(\Delta, h) = J \sum_{n=1}^{N} \left[ \left( S_n^x S_{n+1}^x + S_n^y S_{n+1}^y \right) + \Delta S_n^z S_{n+1}^z \right] - h \sum_{n=1}^{N} S_n^z \quad \quad (S1)$$

in which the Ising anisotropy $\Delta \equiv \cosh \eta > 1$, and $h$ is related to the external magnetic field $B$ by $h \equiv g\mu_B B$. Taking the fully polarized state $|\dots \rightarrow\rightarrow\rightarrow \cdots\rangle$ as the reference state, one can divide the states into subspaces according to the number of flipped spins $r$ or equivalently the total spin in $z$-direction $S_T^z = N/2 - r$. In the corresponding subspace, the trial wavefunction is taken as

$$|\Psi(\{k_j\})\rangle = \sum_{1 \le n_1 \le \cdots \le n_r \le N} a(\{k_j\}, \{n_j\}) \prod_{j=1}^{r} S_{n_j}^{-} |\dots \uparrow\uparrow\uparrow \cdots\rangle \quad \quad (S2)$$

which is parametrized by $r$ momenta $k_j$ ($1 \le j \le r$). In the Bethe ansatz formalism, the coefficients are related to the momenta through the expression (*1,2,3,22,23,26,33*)

$$a(\{k_j\}, \{n_j\}) = \sum_{P} \exp\left( i \sum_{j=1}^{r} k_{P_j} n_j + \frac{i}{2} \sum_{i<j} \theta_{P_i P_j} \right), \quad \quad (S3)$$

in which $P$ is a permutation of the labels $\{1, 2, \dots, r\}$ and the summation is over all permutations; the rapidities $\{\lambda_j\}_{1 \le j \le r}$ are related to the momenta by

$$e^{ik_j} \equiv \frac{\sin\left(\lambda_j + \frac{i\eta}{2}\right)}{\sin\left(\lambda_j - \frac{i\eta}{2}\right)}; \tag{S4}$$

and the phase shift $\theta_{P_iP_j} \equiv \theta_2(\lambda_{P_i} - \lambda_{P_j})$ is defined by taking $n = 2$ in the following expression

$$\theta_n(\lambda) \equiv 2\arctan\left(\frac{\tan\lambda}{\tanh\left(\frac{n\eta}{2}\right)}\right) + 2\pi\left\lfloor\frac{\mathrm{Re}(\lambda)}{\pi} + \frac{1}{2}\right\rfloor \tag{S5}$$

in which $\lfloor A \rfloor$ denotes the floor function giving the greatest integer not larger than $A$.

The energy and momentum of the corresponding state are determined from the momenta $\{k_j\}$ by

$$E = \sum_{j=1}^{r} \epsilon(k_j) = J\sum_{j=1}^{r}(\cos k_j - \Delta) \tag{S6}$$

and

$$k = \sum_{j=1}^{r} k_j, \tag{S7}$$

respectively.

The rapidities, hence the wavefunctions, can be obtained by solving the following Bethe ansatz equations

$$N\theta_1(\lambda_j) = 2\pi I_j + \sum_{l=1}^{r}\theta_2(\lambda_j - \lambda_l), \quad j = 1, \ldots, r, \tag{S8}$$

in which $\theta_1(\lambda)$ and $\theta_2(\lambda)$ are defined in Eq. (S5). The Bethe quantum numbers $\{I_j\}_{1\leq j\leq r}$ take integer values when $r$ is odd, and half-integer values when $r$ is even (see following sections). A state is called real Bethe eigenstate if all the rapidities $\lambda_j$'s are real, and is called string state if there is a complex-valued $\lambda_j$.

### $n$-pair psinon-(anti)psinons (real Bethe eigenstates) and their Bethe quantum numbers

In the subspace with $r$ flipped spins, i.e. $S_T^z = N/2 - r$, the Bethe quantum numbers of the ground state are given by

$$I_j = -\frac{r+1}{2} + j, \quad 1 \leq j \leq r. \tag{S9}$$

A schematic plot of the distribution of ground state Bethe quantum numbers is shown in Fig. S4a for $N = 32$ and $S_T^z = 8$.

The low-lying excitations of real Bethe states can be classified according to either one of the two different patterns that have purely real rapidities $\{\lambda_j\}_{1\leq j\leq r}$ (19): $n$-pair psinon-psinons ($n\psi\psi$) and $n$-pair psinon-antipsinons ($n\psi\psi^*$). The Bethe quantum numbers of an $n\psi\psi$ state can be produced by first extending the left (right) edge of the ground state Bethe quantum numbers further to the left (right) by $n$, then removing $2n$ numbers within the extended range. Those of an $n\psi\psi^*$ state can be obtained by removing $n$ numbers in the ground state range and put them outside the range. Pictorial illustration of the two situations is shown in Fig. S4b and S4c for $1\psi\psi$ and $1\psi\psi^*$, respecitvely.

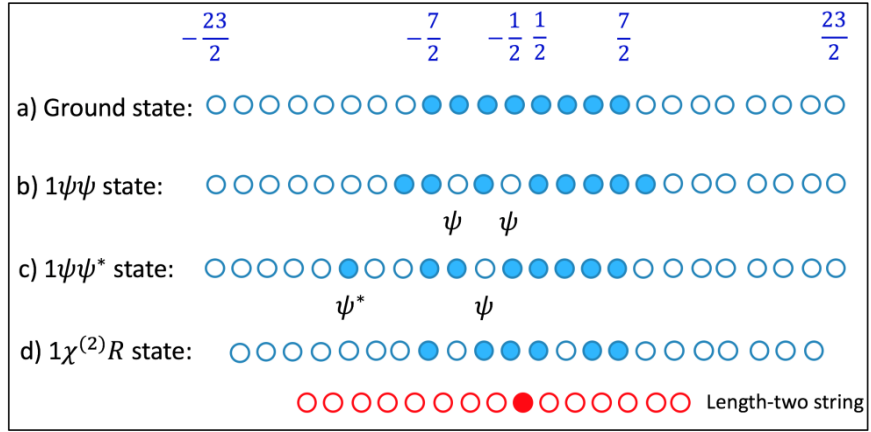

**Fig. S4**. Schematic plots of patterns of Bethe quantum numbers for a) the ground state, b) one-pair psinon-psinon state $1\psi\psi$, c) one-pair psinon-antipsinon state $1\psi\psi^*$, and d) length-two string state $1\chi^{(2)}R$. The system size is taken as $N = 32$, and the magnetization is $S_T^z = 8$.

### String states and their Bethe quantum numbers

The excited states containing complex rapidities are dubbed as string states (2,3,22,23). For a state of a single $n$-string, there are $n$ corresponding complex rapidities

$$\lambda_j^n = \lambda^{(n)} + i(n + 1 - 2j)\frac{\eta}{2} + \delta_j^n, \qquad 1 \leq j \leq n, \tag{S10}$$

in which $\lambda^{(n)}$ is real and referred to as string center, $j$ indexes the rapidities inside the string, and $\delta_j^n$ parameterizes the string deviation.

In string ansatz, the string deviations are neglected. For a general string state of $r$ down-spins, consisting of $r_n$ $n$-strings with $\sum_n nr_n = r$, the string centers are determined by the reduced Bethe-Gaudin-Takahashi equations as (3,33)

$$N\theta_n(\lambda_\alpha^{(n)}) = 2\pi I_\alpha^{(n)} + \sum_{(m\beta)\neq(n\alpha)} \Theta_{nm}(\lambda_\alpha^{(n)} - \lambda_\beta^{(m)}), \quad 1 \leq \alpha \leq r_n, 1 \leq \beta \leq r_m, \tag{S11}$$

where

$$\Theta_{nm} = (1 - \delta_{nm})\theta_{|n-m|} + 2\theta_{|n-m|+2} + \cdots + 2\theta_{n+m-2} + \theta_{|n+m|}, \tag{S12}$$

and $I_\alpha^{(n)}$ are the reduced Bethe quantum numbers of the corresponding strings.

Combining the patterns of psinons/antipsinons with the strings, one can produce excitations like $1\chi^{(n)}R$, where $R$ can be either $n\psi\psi$ or $n\psi\psi^*$ with the real rapidities. A schematic plot of such possibility is shown in Fig. S4d.

### Theoretical magnetization versus magnetic field

We define $H_0$ as the Hamiltonian without magnetic field,

$$H_0(\Delta) = J \sum_{n=1}^{N} \left[ \left( S_n^x S_{n+1}^x + S_n^y S_{n+1}^y \right) + \Delta S_n^z S_{n+1}^z \right], \tag{S13}$$

and $m$ as magnetization per site (20),

$$m = \frac{1}{N} \langle G | S_T^z | G \rangle, \tag{S14}$$

where $|G\rangle$ is the ground state. Then magnetic field $h$ and magnetization $m$ are related by the Legendre transform as

$$h = \frac{\partial e_0}{\partial m}, \tag{S15}$$

with

$$e_0 = \frac{1}{N} \langle G | H_0 | G \rangle. \tag{S16}$$

### Transverse dynamic structure factor

The dynamic structure factors (DSFs) relevant to the experiment are

$$S^{+-}(q,\omega) = \pi \sum_\mu \left| \langle \mu | S_q^- | G \rangle \right|^2 \delta(\omega - E_\mu + E_G), \tag{S17}$$

and

$$S^{-+}(q,\omega) = \pi \sum_\mu \left| \langle \mu | S_q^+ | G \rangle \right|^2 \delta(\omega - E_\mu + E_G), \tag{S18}$$

in which $S_q^\pm$ is the Fourier transform of $S_n^\pm$, i.e. $S_q^\pm = \frac{1}{\sqrt{N}} \sum_n e^{iqn} S_n^\pm$, and $E_G$ and $E_\mu$ are the energies of the ground state $|G\rangle$ and the excited state $|\mu\rangle$, respectively.

The calculations are carried out for the Ising anisotropy $\Delta = 2$ and system size $N = 200$ using the determinant formulas in algebraic Bethe ansatz formalism (26). Fig. S5 displays the results of $S^{+-}(q,\omega)$ and $S^{-+}(q,\omega)$ at a representative magnetization $2m = 0.4$. $S^{-+}(q,\omega)$ is dominated by gapless continua that are formed by $1\psi\psi$ and $2\psi\psi$ excitations, whereas in $S^{+-}(q,\omega)$, there are several well-separated dynamical branches. For $S^{+-}(q,\omega)$, the lowest-lying gapless continua are formed by $1\psi\psi^*$ and $2\psi\psi^*$ excitations. The corresponding spectral weights are located mainly in the energy range $\hbar\omega < 3J$. The higher-energy separated continua correspond to 2-string and 3-string excitations that are found in the spectral range $\hbar\omega > 3J$ and $\hbar\omega > 5J$, respectively.

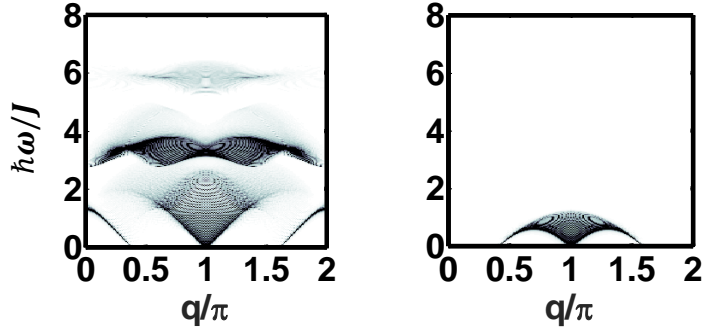

**Fig. S5**. Dynamic structure factors $S^{+-}(q,\omega)$ (left) and $S^{-+}(q,\omega)$ (right) as a function of energy $\hbar\omega/J$ (vertical axis) and momentum $q/\pi$ (horizontal axis) for $2m = 0.4$, $N = 200$. The gapless continua are formed by real Bethe eigenstates (psinon-antipsinon pairs in $S^{+-}$ and psinon-psinon pairs in $S^{-+}$). For $S^{+-}$, the higher-energy continua correspond to excitations of 2-string ($\hbar\omega > 3J$) and 3-string ($\hbar\omega > 5J$), respectively.

### The momentum integrated sum rule

By rescaling $S^{\pm}$ to $\frac{1}{\sqrt{2}}S^{\pm}$, the momentum integrated sum rule can be expressed as

$$R_{a\bar{a}} = \frac{1}{N}\sum_q \int_0^\infty \frac{d\omega}{2\pi} S^{a,\bar{a}}(q,\omega) = \frac{1}{4} + \frac{m}{2}c_a, \quad (S19)$$

where $c_a = \pm 1$, for $a = \pm$, respectively, for the transverse dynamic structure factors.

To evaluate the saturation levels of the sum rules, we define the ratio

$$\nu_{a\bar{a}} = \frac{R'_{a\bar{a}}}{R_{a\bar{a}}}, \quad (S20)$$

where $R'_{a\bar{a}}$ is calculated from a partial summation over the selected excitations.

### The momentum integrated ratios of transverse DSFs for psinon-(anti)psinons and string excitations

The momentum integrated ratios are displayed in Fig. S6(a) and Fig. S6(b), respectively, for $S^{+-}$ and $S^{-+}$. The calculations are carried out for $2m$ varied from 0.1 to 0.9 with step 0.1. The good saturations of sum rules (above 87%) for both $S^{+-}$ and $S^{-+}$ over the entire range of magnetization indicate that most of the spectral weights are accounted in the calculations. Particularly for $S^{+-}$, the string excitations especially 3-string states become progressively important when magnetization is lowered. Hence, the string states play a predominant role in the dynamical properties in the low magnetization region.

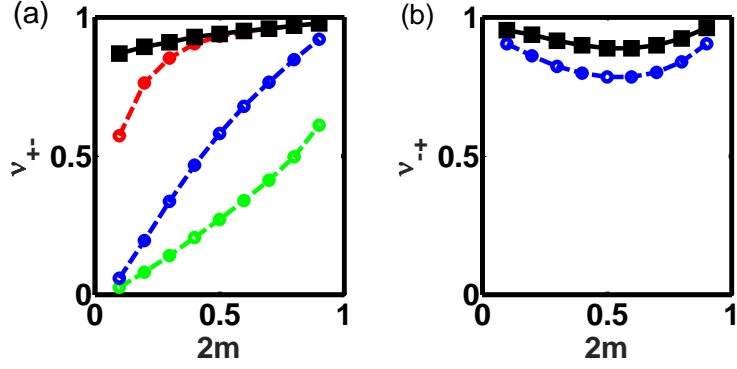

**Fig. S6**. The momentum integrated ratios (a) $\nu_{+-}$ and (b) $\nu_{-+}$ (Eqs. S19, S20) from $S^{+-}$ and $S^{-+}$ channels (Eqs. S17, S18), respectively, as a function of magnetization $2m$. In (a), the green line is the $1\psi\psi^*$ contribution. The blue, red and the black lines are augmented by progressively taking into account the $2\psi\psi^*$, 2-string, and 3-string contributions, respectively. In (b), the blue and black lines represent the $1\psi\psi$ and $1\psi\psi + 2\psi\psi$ contributions.

## Comparison between theory and experiment

With the four-fold screw axis (Fig. S1A), the Brillouin zone is folded by a factor of four in SrCo$_2$V$_2$O$_8$, thus besides at $q = 0$, the excitations at $q = \pi$ and $q = \pi/2$ also should be considered for comparison with the THz experiments. Based on the results shown in Fig. S5, in the following we present a detailed analysis of the magnetic-field-dependent properties of the various excitations for these $q$ momenta.

Figures S7 – S10 show the dynamic structure factors (DSFs) for psinon-psinon, psinon-antipsinon, 2-string, and 3-string, respectively, as functions of energy for the magnetizations $2m = 0.1$ to $0.9$ and $q = 0, \pi/2$ and $\pi$.

As shown in Fig. S7, for the psinon-psinon excitations the DSF spectra exhibit very sharp peaks and the peak positions evidently shift with magnetic field (or magnetization) only for $q = \pi/2$. In contrast, at $q = 0$ and $q = \pi$ the excitations form continua in the whole spectral range $\sim 2J$, and one cannot define peak positions, or the peak positions are almost independent on magnetic field. Thus, one can clearly distinguish the psinon-psinon excitations at $q = \pi/2$ from those at $q = 0$ and $q = \pi$. Moreover, the softening of the $q = \pi/2$ modes on increasing magnetic fields is in clear contrast to the hardening of the other types of excitations (see Fig. 4). This is because psinon-psinon excitations correspond to flipping one spin parallel to the magnetic field. Indeed, the psinon-psinon excitations at $q = \pi/2$ are observed by our THz spectroscopy (see Fig. 4).

The psinon-antipsinon excitations exhibit very sharp peaks at $q = 0$ and $q = \pi/2$ but form continua at $q = \pi$ (see Fig. S8). At $q = 0$, one can clearly see the peaks in the whole field range (Fig. S8A), while for $q = \pi/2$, only above half-magnetization saturation, peaks can be well resolved (Fig. S8B). For both of the modes, the peak positions shift to higher energy with increasing magnetic fields. The psinon-antipsinon excitations at $q = 0$ are observed experimentally for all the magnetic fields above $B_c = 4$ T (see Fig. 4), while those at $q = \pi/2$, which appear above $B_{hs} = 25$ T and at low energies, are yet to be found. The extended continua at $q = \pi$ are strongly overlapping for different magnetic fields, and thus they cannot be resolved by the present magneto-optical spectroscopy.

Figure S9 shows the DSFs of 2-string excitations. At $q = \pi$ the spectra exhibit sharp peaks with well-defined peak positions, while at $q = 0$ the peaks are relatively broad. With increasing magnetic fields, these two modes are evidently hardening and thus can be well identified. In contrast, the continua at $q = \pi/2$ overlap with each other for different magnetic fields. Therefore, only at $q = \pi$ and $q = 0$ the 2-string excitations can be experimentally observed (see Fig. 4).

The 3-string excitations form extended continua at $q = 0$ and $q = \pi$, which overlap for the different magnetic fields, as shown in Fig. S10. In contrast, at $q = \pi/2$ the excitation spectra exhibit well-defined peaks with peak positions shifting clearly to higher energy with increasing magnetic field. This mode can be clearly identified experimentally (see Fig. 4).

To summarize, only the excitations with well-defined peak positions which shift evidently in magnetic field can be resolved, and as have been observed for $R_0$, $R_{\pi/2}$, $\chi_0^{(2)}$, and $\chi_\pi^{(2)}$, and $\chi_{\pi/2}^{(3)}$ by our magneto-optic THz spectroscopy (see Fig. 4). In contrast, the other excitations form continua in a broad spectral range and cannot be easily resolved by the THz spectroscopy.

The theoretical peak positions of the excitations $R_0$, $R_{\pi/2}$, $\chi_0^{(2)}$, and $\chi_\pi^{(2)}$, and $\chi_{\pi/2}^{(3)}$ are plotted as functions of magnetic field and compared to the experimental results. We fix the exchange interaction and Ising anisotropy to the previously determined values for SrCo$_2$V$_2$O$_8$, i.e. $J = 3.55$ meV and $\Delta = 2$ (27), and with the g-factor $g_\parallel = 6.2$ all the five experimentally observed modes can be very well described by the theory, as presented in Fig. 4.

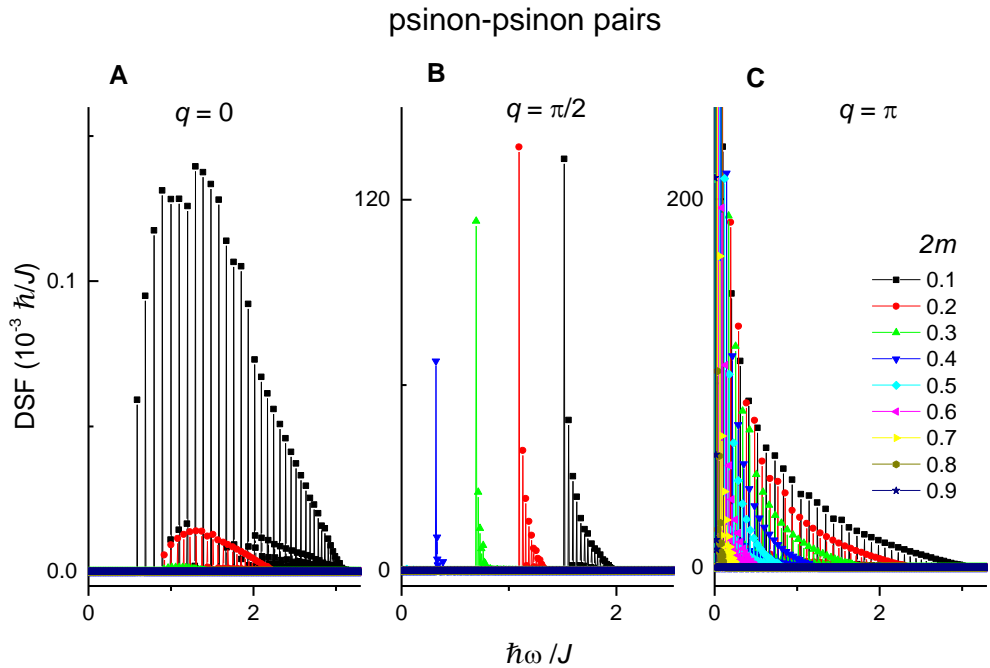

**Fig. S7.** Dynamic structure factor of psinon-psinon pairs as a function of energy for $2m = 0.1$ to $0.9$ at $q = 0$, $\pi/2$, and $\pi$.

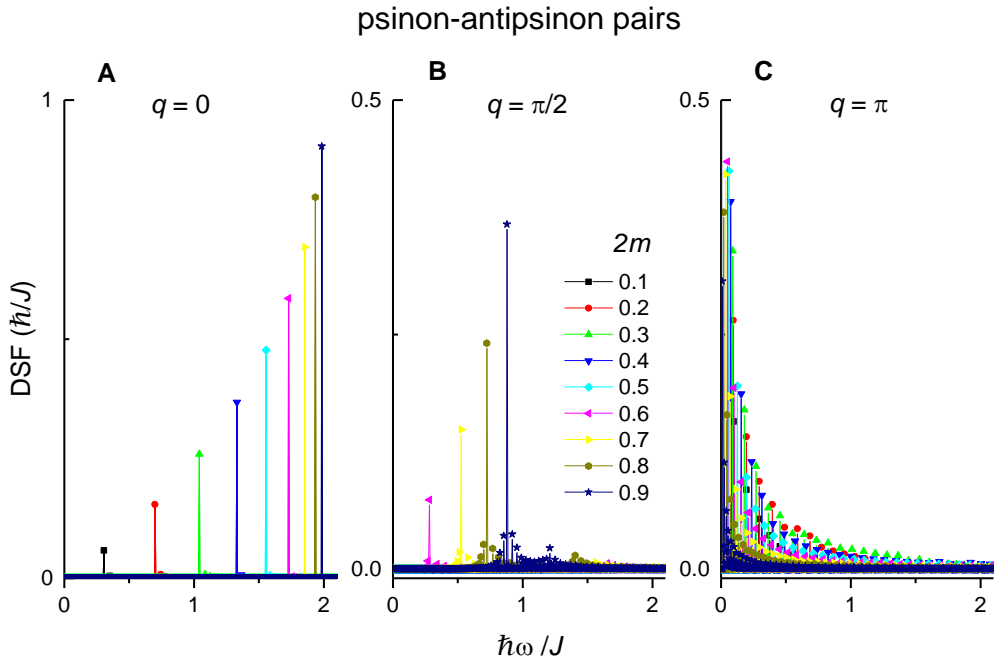

**Fig. S8.** Dynamic structure factor of psinon-antipsinon pairs as a function of energy for $2m = 0.1$ to $0.9$ at $q = 0, \pi/2$, and $\pi$.

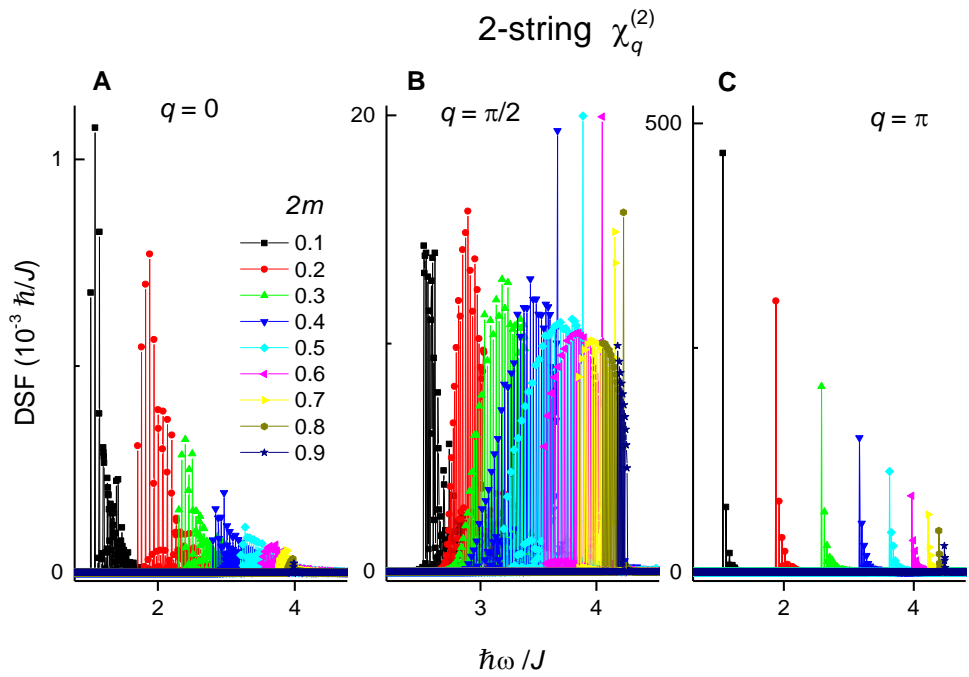

**Fig. S9.** Dynamic structure factor of 2-string as a function of energy for $2m = 0.1$ to $0.9$ at $q = 0, \pi/2$, and $\pi$.

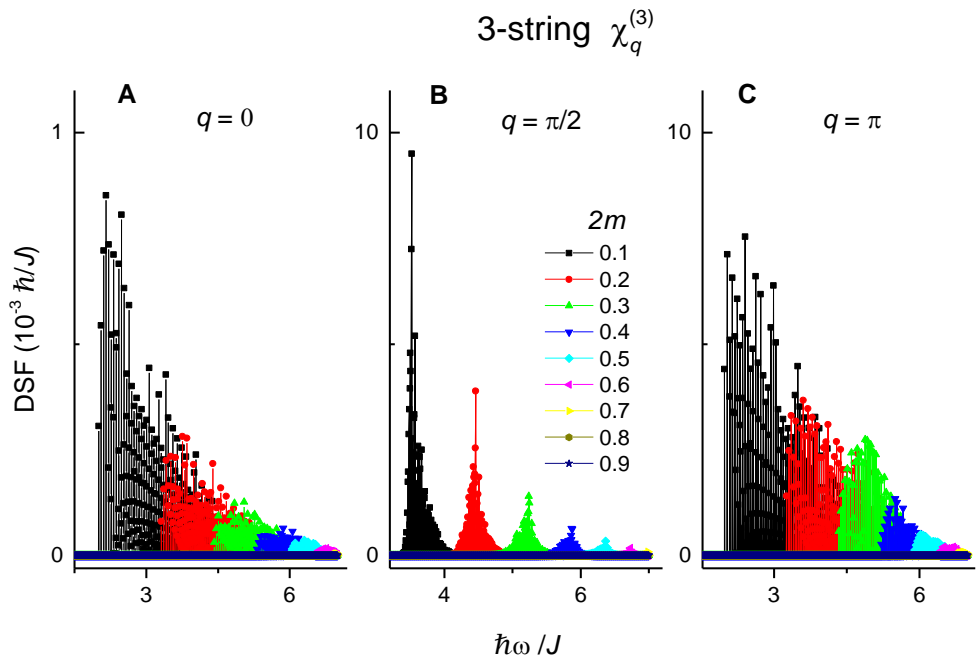

**Fig. S10.** Dynamic structure factor of 3-string as a function of energy for $2m = 0.1$ to $0.9$ at $q = 0, \pi/2$, and $\pi$.

---

[31] B. Grenier *et al.*, *Phys. Rev. Lett.* **114**, 017201 (2015).
[32] H. Shiba *et al.*, *J. Phys. Soc. Jpn.* **72**, 2326–2333 (2003).
[33] M. Takahashi, *Thermodynamics of One-Dimensional Solvable Models* (Cambridge University Press, 2005)